\documentclass[]{spie}  %>>> use for US letter paper
\usepackage{aas_macros}
\usepackage{amsmath,amsfonts,amssymb}
\usepackage{rotating}
\usepackage{graphicx}
\usepackage[colorlinks=true, allcolors=blue]{hyperref}
\usepackage{threeparttable}

\title{Superfluid-tight cryogenic receiver with continuous sub-Kelvin cooling for EXCLAIM}

\author[a,b]{Sumit Dahal}
\author[c]{Peter A. R. Ade}
\author[a]{Christopher J. Anderson}
\author[a]{Alyssa Barlis}
\author[a]{Emily M. Barrentine}
\author[d]{Jeffrey W. Beeman}
\author[a]{Nicholas G. Bellis}
\author[e]{Alberto D. Bolatto}
\author[f]{Victoria Braianova}
\author[g]{Patrick C. Breysse}
\author[a]{Berhanu T. Bulcha}
\author[a]{Giuseppe Cataldo}
\author[a]{Felipe A. Colazo}
\author[a]{Lee-Roger Chevres-Fernandez}
\author[a]{Chullhee Cho}
\author[a]{Danny S. Chmaytelli}
\author[a]{Jake A. Connors}
\author[a]{Nicholas P. Costen}
\author[a]{Paul W. Cursey}
\author[a]{Negar Ehsan}
\author[a]{Thomas M. Essinger-Hileman}
\author[a]{Jason Glenn}
\author[h]{Joseph E. Golec}
\author[a]{James P. Hays-Wehle}
\author[a]{Larry A. Hess}
\author[a]{Amir E. Jahromi}
\author[a]{Trevian Jenkins}
\author[a]{Mark O. Kimball}
\author[a]{Alan J. Kogut}
\author[f]{Samuel H. Kramer}
\author[a]{Nicole Leung}
\author[a]{Luke N. Lowe}
\author[i]{Philip D. Mauskopf}
\author[h]{Jeffrey J. McMahon}
\author[j,a]{Vilem Mikula}
\author[a]{Mona Mirzaei}
\author[a]{Samuel H. Moseley}
\author[a]{Jonas W. Mugge-Durum}
\author[a]{Jacob Nellis}
\author[a]{Omid Noroozian}
\author[i]{Kate Okun}
\author[a]{Trevor Oxholm}
\author[a]{Tatsat Parekh}
\author[k]{Ue-Li Pen}
\author[g]{Anthony R. Pullen}
\author[a]{Maryam Rahmani}
\author[a]{Mathias M. Ramirez}
\author[i]{Cody Roberson}
\author[a]{Samelys Rodriguez}
\author[a]{Florian Roselli}
\author[a]{Deepak Sapkota}
\author[a]{Konrad Shire}
\author[f]{Gage L. Siebert}
\author[f]{Faizah Siddique}
\author[i]{Adrian K. Sinclair}
\author[l]{Rachel S. Somerville}
\author[i]{Ryan Stephenson}
\author[a]{Thomas R. Stevenson}
\author[a]{Eric R. Switzer}
\author[a]{Jared Termini}
\author[f]{Peter T. Timbie}
\author[a]{Justin Trenkamp}
\author[c]{Carole E. Tucker}
\author[m]{Elijah Visbal}
\author[a,e]{Carolyn G. Volpert}
\author[a]{Joseph Watson}
\author[i]{Eric Weeks}
\author[a]{Edward J. Wollack}
\author[l]{Shengqi Yang}
\author[a]{Aaron Yung}

\affil[a]{NASA Goddard Space Flight Center, Greenbelt, MD, USA}
\affil[b]{Johns Hopkins University, Baltimore, MD, USA}
\affil[c]{Cardiff University, Cardiff, Wales, UK}
\affil[d]{Lawrence Berkeley National Lab, Berkeley, CA, USA}
\affil[e]{University of Maryland, College Park, MD, USA}
\affil[f]{University of Wisconsin-Madison, Madison, WI, USA}
\affil[g]{New York University, New York, NY, USA}
\affil[h]{University of Chicago, Chicago, IL, USA}
\affil[i]{Arizona State University, Tempe, AZ, USA}
\affil[j]{Science Systems and Applications, Inc., Lanham, MD, USA}
\affil[k]{Canadian Institute for Theoretical Astrophysics, Toronto, ON, Canada}
\affil[l]{Flatiron Institute, New York, NY, USA}
\affil[m]{University of Toledo, Toledo, OH, USA}

\authorinfo{Further author information: (Send correspondence to S.D.)\\S.D.: E-mail: sumit.dahal@nasa.gov}

\begin{document} 
\maketitle

\begin{abstract}
The EXperiment for Cryogenic Large-Aperture Intensity Mapping (EXCLAIM) is a balloon-borne telescope designed to survey star formation over cosmological time scales using intensity mapping in the 420 – 540 GHz frequency range. EXCLAIM uses a fully cryogenic telescope coupled to six on-chip spectrometers featuring kinetic inductance detectors (KIDs) to achieve high sensitivity, allowing for fast integration in dark atmospheric windows. The telescope receiver is cooled to $\approx$ 1.7 K by immersion in a superfluid helium bath and enclosed in a superfluid-tight shell with a meta-material anti-reflection coated silicon window. In addition to the optics and the spectrometer package, the receiver contains the magnetic shielding, the cryogenic segment of the spectrometer readout, and the sub-Kelvin cooling system. A three-stage continuous adiabatic demagnetization refrigerator (CADR) keeps the detectors at 100~mK while a $^4$He sorption cooler provides a 900 mK thermal intercept for mechanical suspensions and coaxial cables. We present the design of the EXCLAIM receiver and report on the flight-like testing of major receiver components, including the superfluid-tight receiver window and the sub-Kelvin coolers.
\end{abstract}

% Include a list of keywords after the abstract 
\keywords{intensity mapping, EXCLAIM, superfluid, continuous adiabatic demagnetization refrigerator}

\section{INTRODUCTION}
\label{sec:intro}
Conventional galaxy surveys create large catalogs of galaxies to study the formation and evolution of large-scale structures in the universe. These surveys are often biased to detect only the brightest galaxies and have small survey areas on the sky, limiting their ability to capture a complete picture of galaxy populations in the cosmological context. The EXperiment for Cryogenic Large-Aperture Intensity Mapping (EXCLAIM)\cite{switzer2021} employs an emerging technique of line intensity mapping \cite{bernal2022} that surveys the unresolved, integral surface brightness of redshifted line emission from galaxies. This approach measures the cumulative emission of all sources over large volumes, allowing a blind, complete census. In particular, EXCLAIM aims to map the redshifted emission of carbon monoxide (CO) and singly-ionized carbon ([CII]) in cross-correlation with spectroscopic galaxy surveys in windows over the $0 < z < 3.5$ redshift range. As this range encompasses the period of ``cosmic high noon'' when the rate of star formation peaked \cite{madau2014}, EXCLAIM's measurements will be crucial in refining galaxy evolution models during this critical period.

EXCLAIM is a balloon-borne telescope designed to map diffuse emission from 420 to 540 GHz (714 to 555~$\mathrm{\mu}$m) with a spectral resolving power of $R$ = 512, covering several CO rotational lines ($\nu_\mathrm{CO,J}$ = 115J GHz for J = 4 -- 7) in galaxies with redshifts $z < 1$ and [CII] ($\nu_\mathrm{[CII]}$ = 1.889 THz) over redshifts $2.5 < z < 3.5$ \cite{Pullen2023}. EXCLAIM's primary extragalactic science comes from a $\sim$ 300~deg$^2$ survey along the celestial equator in cross-correlation with galaxy and quasar catalogs in the overlapping Stripe-82 region mapped by multiple surveys, particularly the Baryon Oscillation Spectroscopic Survey (BOSS)\cite{alam2021}. The EXCLAIM survey also includes several $\sim$100 deg$^2$ regions on the Galactic plane to study CO (J=4--3) and neutral carbon ([CI]) emission as tracers for star formation and molecular gas. For further details on the EXCLAIM survey and science forecasts, refer to Refs. \citenum{switzer2021} and \citenum{Pullen2023}.

Based on the Absolute Radiometer for Cosmology, Astrophysics, and Diffuse Emission II (ARCADE II)\cite{singal2011} and the Primordial Inflation Polarization ExploreR (PIPER)\cite{gandilo2016} heritage, EXCLAIM employs a fully cryogenic telescope housed in an open 3000 L liquid helium (LHe) bucket dewar with superfluid fountain effect pumps\cite{kogut2021} that cool the optics to $<$5 K and maintain the receiver cryostat at $\approx$1.7 K. Inside the receiver, six integrated spectrometers\cite{barrentine2016, cataldo2019} coupled to kinetic inductance detectors (KIDs) provide the R = 512 spectral resolving power over the EXCLAIM frequency band. At target balloon float altitudes above 27 km, low total atmospheric column depth and pressure broadening cause the atmospheric emission to resolve into narrow lines, allowing the high-resolution integrated spectrometers to access low-background windows between the lines. The all-cryogenic instrument design enables EXCLAIM to fully utilize the dark atmospheric windows in the stratosphere by allowing access to spectral channels $\approx50\times$ darker\cite{switzer2021} than those with ambient temperature optics. This high sensitivity drastically increases mapping speed, enabling EXCLAIM to achieve its science goals with a single-day conventional balloon flight that would otherwise take weeks.

\begin{sidewaysfigure}
\begin{center}
\includegraphics[scale=0.68, trim=70 0 50 0]{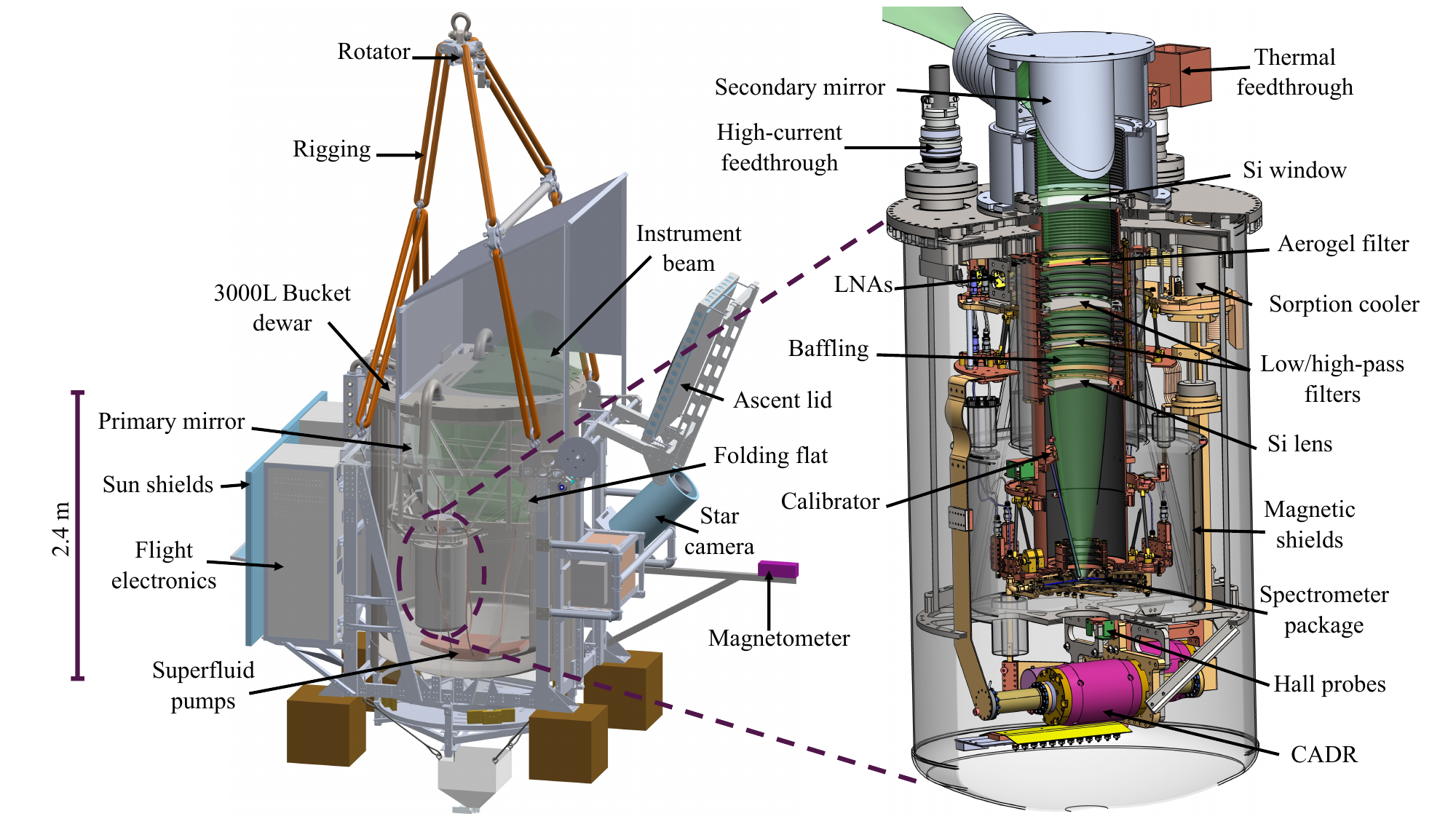}
\end{center}
\caption{ \label{fig:exclaim_CAD} \textit{Left:} CAD model of the EXCLAIM gondola showing the fully cryogenic telescope housed in an open helium bucket dewar. The green region shows the instrument beam to the sky. \textit{Right:} Zoomed-in sectional view of the receiver cryostat highlighting the major optical and thermal components. All receiver interfaces must remain superfluid tight.}
\end{sidewaysfigure}

At float, the EXCLAIM receiver sits in a superfluid helium bath, and the receiver core is initially cooled to $\approx$~1.7~K via a thermal feedthrough (see Figure \ref{fig:exclaim_CAD}). All receiver interfaces must therefore remain superfluid tight to prevent superfluid helium from entering the receiver volume, where superfluid films or gas could hinder sub-Kelvin operation. A three-stage continuous adiabatic demagnetization refrigerator (CADR), backed by a $^4$He sorption cooler, maintains the on-chip spectrometers at their operating temperature of 100 mK during the flight. Here we describe the design and testing of the EXCLAIM receiver that provides superfluid-tight enclosure for continuous sub-Kelvin operation of the sorption-cooler-backed CADR. As the availability of LHe limits the test time for the integrated receiver, we report on the independent testing and qualification of the major receiver components before integrating the receiver core.

This paper is organized as follows. We provide a general overview of the EXCLAIM instrument in Section~\ref{sec:instrument}, and focus on the receiver design in Section~\ref{sec:receiver}. In Section~\ref{sec:test}, we report on the lab testing of major receiver components in flight-like conditions. Finally, Section \ref{sec:summary} summarizes the current status and path towards the first flight from Fort Sumner, New Mexico, planned for September 2025.

\section{Instrument Overview}
\label{sec:instrument}
As shown in Figure \ref{fig:exclaim_CAD}, EXCLAIM employs a fully cryogenic telescope housed in a 3000 L LHe bucket dewar with a 2.0\,m deep and 1.5\,m diameter interior. Since the dewar size drives the overall mass of the gondola, it was chosen to be within a reasonable margin of the total balloon payload mass limit of 3400 kg. During ground operations and ascent, a lid covers the dewar to insulate the instrument, reduce boiloff, and limit the atmosphere from freezing on the optics. The lid is kept open for the science operation at float, letting the boiloff gas keep the optics dry and clean and eliminating the need for an ambient temperature window. EXCLAIM is expected to have a similar boiloff rate of 110 L/h when the superfluid pumps are operational as observed in PIPER \cite{kogut2021}. When launched with a maximum practical helium load of $\sim$ 2600 L, this boiloff rate allows for $\gtrsim$ 19 hours of cryogenic hold time at float, longer than the planned baseline hold time of $\gtrsim$~12~h for the EXCLAIM sub-Kelvin operation.

With the telescope boresight fixed at 45$^\circ$ elevation, EXCLAIM uses a 90-cm parabolic primary mirror and \mbox{10-cm} parabolic secondary mirror in an off-axis Gregorian configuration to produce a collimated beam that couples to the receiver. A 30-cm folding flat mirror redirects the rays from the primary to the secondary mirror so that the telescope can fit within the dewar. The telescope optics, along with a 10-cm silicon lens inside the receiver (see Section \ref{sec:optics}), provides 4.2$^\prime$ full width at half maximum (FWHM) resolution in the center of the EXCLAIM band at 480 GHz over a 22.5$^\prime$ field of view \cite{essingerhileman2020}. This angular resolution is sufficient to produce a survey that covers spatial scales from the linear regime ($k \lesssim 0.1~h\mathrm{Mpc}^{-1}$) up to scales where shot noise dominates ($k \gtrsim 5~h\mathrm{Mpc}^{-1}$) in the line intensity signal \cite{cataldo2020, bernal2019}. All the mirrors were machined from monolithic aluminum at the Johns Hopkins Applied Physics Laboratory. The fabricated mirrors meet the 20 $\mu$m root mean square (RMS) surface figure requirement and have been delivered to NASA Goddard for installation and alignment into the telescope frame. See Ref. \citenum{essingerhileman2020} for further details on the telescope optics.

The telescope optics will be operated at $<$ 5 K to ensure low optical loading on each spectrometer channel ($\sim$~0.1~fW defined at the receiver cold stop) to ensure near background-limited performance across the band, enabling access to dark regions between atmospheric lines. The rapid helium boiloff is sufficient to maintain the telescope at the bath temperature during ascent. At float altitudes of $\gtrsim$ 27 km, the $\lesssim$~1~kPa ambient pressure decreases the boiling point of helium to below 1.7 K, lower than the 2.2 K superfluid transition temperature \cite{kogut2021}. Once the float altitude is reached and the boiloff rate decreases, the superfluid fountain effect pumps at the bottom of the dewar (see Figure \ref{fig:exclaim_CAD}) are turned ON. These pumps spray superfluid helium onto each optical surface to maintain $<$ 5 K temperature during the science operation. The design and performance of these superfluid pumps during the two PIPER flights are described in Ref. \citenum{kogut2021}.

\section{Receiver Design}
\label{sec:receiver}
The receiver cryostat is positioned within the telescope frame using a symmetric hexapod of locking turnbuckles such that the folding flat mirror directs the instrument beam onto the secondary mirror mounted on top of the receiver lid as shown in Figure \ref{fig:exclaim_CAD}. The secondary mirror produces a collimated beam that passes into the receiver cryostat through a meta-material anti-reflection (AR) coated silicon vacuum window (Section \ref{sec:optics}). The 44.6-cm diameter and 72.5-cm height stainless steel receiver shell sits partially submerged in LHe, preventing the receiver window from being submerged during science observation. All the receiver interfaces employ superfluid-tight seals to prevent superfluid helium from entering the receiver volume, where superfluid films or gas could hinder sub-Kelvin cooling.

The receiver houses (1) the spectrometer package, (2) the optical filtering, baffling, and lens, (3) the magnetic shielding, (4) the sub-Kelvin cooling system, (5) the cryogenic segment of the spectrometer readout, (6) electrical interfaces to the ambient temperature electronics, and (7) thermal interfaces to the helium bath. In the following subsections, we describe the design of the major optical and thermal components inside the receiver.

\subsection{Optics}
\label{sec:optics}

\begin{figure}[ht]
   \begin{center}
   \begin{tabular}{cc}
   \includegraphics[height=3.5cm]{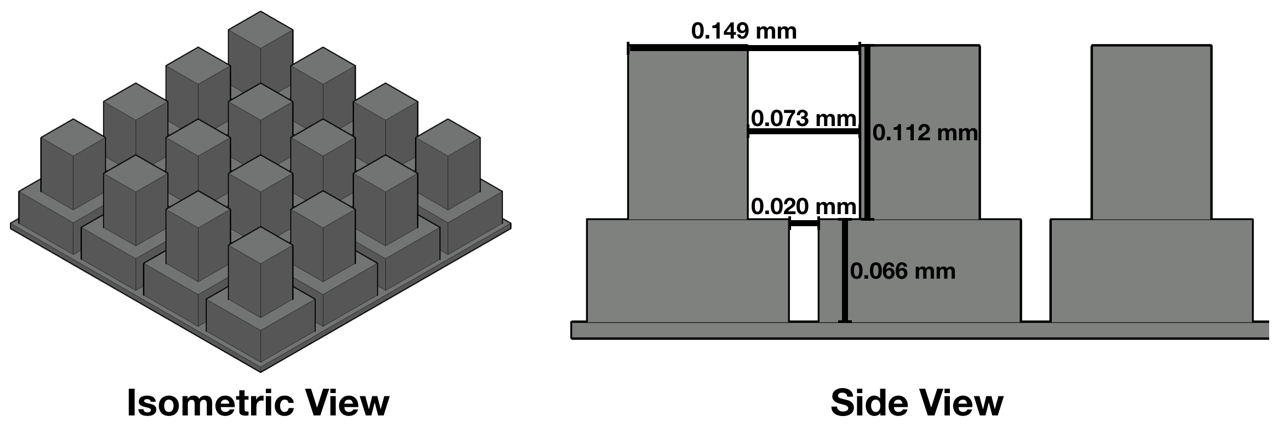}
   \includegraphics[height=4.23cm]{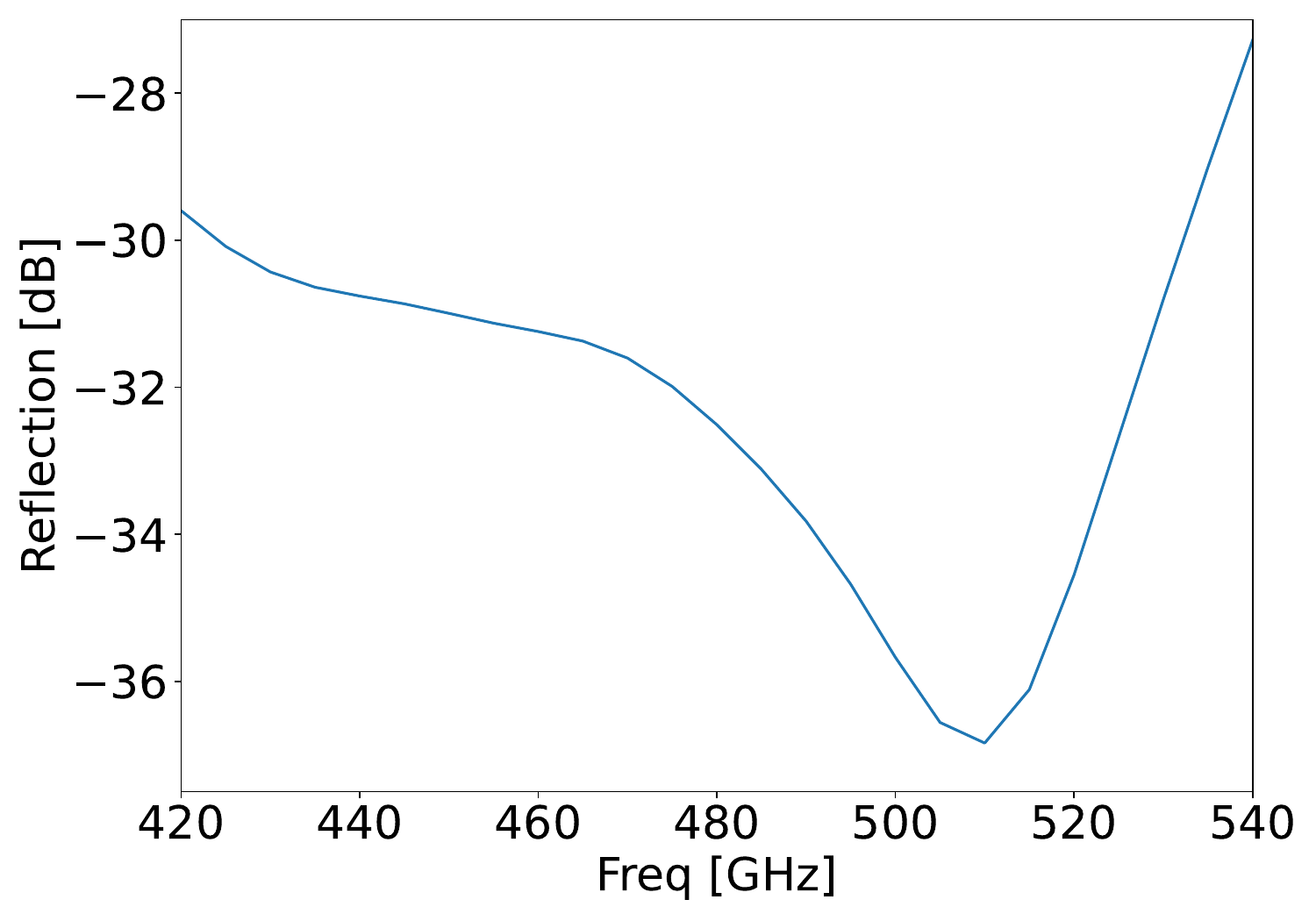}
   \end{tabular}
   \end{center}
   \caption{ \label{fig:AR_coating} The isometric (\textit{left}) and side (\textit{middle}) views of the optimized two-layered anti-reflection coating design for the EXCLAIM silicon window and lens. The \textit{right} plot shows the simulated performance at normal incidence for the design, achieving $< -27$~dB reflection across the band.}
   \end{figure}

The 114-mm diameter open aperture on the receiver lid uses a silicon window of 9 mm thickness, required to support the ambient atmospheric pressure prior to launch. It employs an indium seal to create a superfluid-tight interface with the receiver lid. We demonstrate this superfluid-tight seal in flight-like conditions in Section \ref{sec:window_test}. The meta-material AR coating is implemented through sub-wavelength features cut into the window surface with a custom dicing saw \cite{datta2013}. As shown in Figure \ref{fig:AR_coating}, EXCLAIM uses an optimized two-layer design to achieve low reflection ($< -27$~dB for normal incidence) across the signal band. A 24-cm focal length plano-convex silicon lens (see Figure \ref{fig:exclaim_CAD}) that focuses light onto the spectrometer package also uses the same AR coating design.

In addition to the cold telescope optics, EXCLAIM requires effective infrared (IR) rejection and stray light control to reduce excess loading on the detectors for accessing the dark spectral channels. We use two 10-cm diameter novel IR-blocking filters composed of diamond scattering particles embedded in a polyimide aerogel substrate. The substrate has an ultra-low density (0.1 -- 0.2 g/cm$^{3}$) with a low index of refraction (n $\sim$ 1.1), eliminating the need for AR coating and allowing for high transmission across the band \cite{essingerhileman2020_aerogel}. The size and density of the diamond particles were tuned to produce a low-pass filter with $\sim$~1~THz cutoff \cite{barlis2022}. To reject radiation immediately above and below the EXCLAIM band, we use two heat-pressed stacks of metal-mesh filters on polypropylene film with dielectric spacers \cite{ade2006}. These band-defining high- and low-pass filters and the lens are tilted at alternating angles (3$^{\circ}$ for the lens and 2$^{\circ}$ for the filters with respect to the chief ray) to suppress the formation of cavity modes and optical ghosts by terminating reflections in baffling.

A collection of baffles, blackened with Epotek 377 epoxy loaded with silica and graphite powders \cite{chuss2017}, are strategically placed at multiple places inside the receiver (see Figure \ref{fig:exclaim_CAD}) for stray light control. A cold (1.7 K) optical stop with 7.6-cm diameter aperture placed in between the lens and the spectrometer package truncates the beam at $< -15$~dB across the band. In combination with the blackened baffles in the collimated region, the cold stop provides $< -40$~dB stray light spill onto the warmer elements at the top of the dewar to ensure $<$~0.1~fW excess loading per spectrometer channel. An in-flight calibration source \cite{Pisano2005}, consisting of a sapphire square with an integrated heater thermally isolated from the bath, is located in the volume behind the cold stop. This calibrator emits into the near sidelobes of the spectrometer lenslets for in-situ characterization of spectrometer response, uniformity, and time-varying responsivity \cite{essingerhileman2020}.

\subsection{Integrated Spectrometer}
\label{sec:spectrometer}

The receiver optics focus the incoming light onto six 4-mm diameter hyper-hemispherical silicon lenslets with 126-$\mu$m Parylene-C AR-coating on the focal plane. Each lenslet couples light to an individual spectrometer chip using a dipole slot antenna. The six integrated spectrometers ($\mu$-Spec)\cite{barrentine2016, cataldo2019} incorporate a Rowland grating spectrometer implemented in a parallel plate waveguide on a low-loss single-crystal silicon chip, employing superconducting niobium microstrip planar transmission lines and thin-film aluminum KIDs \cite{volpert2022}. The  $\mu$-Spec design offers several advantages: (1) an order of magnitude reduction in size compared to a free-space grating spectrometer, (2)~lithographic control of all components, (3) high efficiency and resolution due to the low dielectric loss of single-crystal silicon, and (4) high immunity to stray light and crosstalk due to the microstrip architecture and thin dielectric \cite{switzer2021}. Under the 0.16 fW loading expected at the input to the KIDs, we estimate the noise-equivalent power NEP$_\mathrm{det} < 8 \times 10^{-19}~\mathrm{W/\sqrt{Hz}}$, within the sensitivity requirements for the EXCLAIM science mission. While the flight-candidate spectrometer wafers are currently under fabrication, a prototype wafer with $R$ = 64 spectral resolution has been extensively characterized in the lab \cite{rahmani2023}.

The KIDs are read through ambient-temperature electronics based on Xilinx ZCU111 Radio-Frequency System-on-a-Chip (RFSoC) FPGA boards that significantly reduce the size, weight, and power requirements and provide larger instantaneous bandwidth compared to previous generation readout systems \cite{sinclair2020}. To read out the six spectrometer chips, we plan to use three ZCU111 boards, each equipped with two 512-MHz bandwidth readout chains. Each readout chain uses a low-noise amplifier (Low Noise Factory LNC2\_4A\footnote{www.lownoisefactory.com/product/lnf-lnc2\_4a-2}) connected to the spectrometer chip with a pair of 2.19 mm OD stainless steel coaxial cables. The six outbound coaxial cables have a beryllium copper centerline that provides lower attenuation at the expense of higher thermal conduction. A 900~mK stage thermally suspended from the 1.7 K bath by a carbon fiber tube truss provides a thermal intercept for the coaxial cable lines going into the 100~mK spectrometer package. The thermal breaks in the cryogenic readout chains use 2.19 mm OD NbTi coaxial cables.

\subsection{Sorption Cooler}
\label{sec:sorption}
Since the stainless steel receiver shell is a poor thermal conductor, we use a thermal feedthrough (Figure \ref{fig:exclaim_CAD}) to bring a high-purity copper rod into the receiver through a superfluid-tight ceramic seal. The reservoir around the thermal feedthrough outside the receiver lid is supplied by superfluid pumps to maintain it at $\sim$~1.7 K. From the 1.7 K bath, we use a single-stage $^4$He sorption cooler \cite{chase2020} to provide a $\sim$ 900 mK stage for pre-cooling the 100 mK stage with the spectrometer package and thermally intercepting the coaxial cables and mechanical suspensions. The sorption cooler provides a significant margin in flight operation by reducing the heat capacity and thermal loading of the CADR (described in Section \ref{sec:CADR}) and allows testing in unpumped helium bath and cryocooler systems in the lab.

The sorption cooler has a simple operation procedure. Once the cold head gets close to the bath temperature (below the 5.2 K critical liquefaction point of $^4$He), we heat the cooler pump to 45 -- 55 K using a 300 $\Omega$ heater on the pump. This keeps the pump warm enough to prevent helium from being adsorbed by the charcoal inside the pump. As the cold head temperature stabilizes, we cool the pump through a gas-gap heat switch. This allows the charcoal to adsorb gaseous helium, lowering the pressure inside the pump while letting the liquid helium in the cold head to continue cooling down until its vapor pressure is in equilibrium with the internal pressure \cite{chase2020}. We estimate that the sorption fridge with 28 J of cooling energy can sufficiently cool the receiver sub-Kelvin stages from the 1.7 K float conditions and provide significant margins over the 12 hours baseline cold operation \cite{switzer2021}.

\subsection{CADR}
\label{sec:CADR}

\begin{figure}[th]
   \begin{center}
   \begin{tabular}{cc}
   \includegraphics[height=6.4cm]{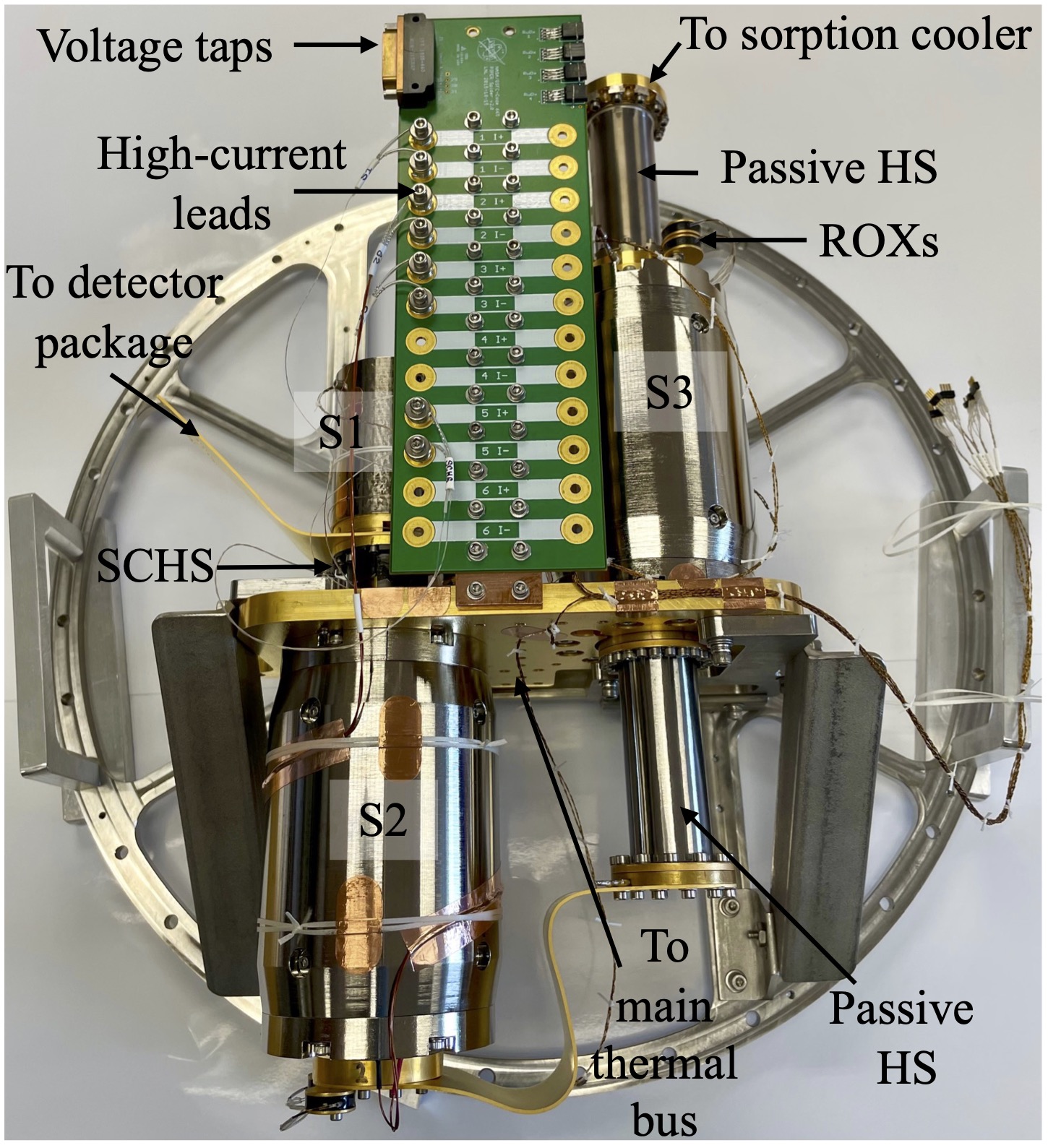}
   \includegraphics[height=6.4cm]{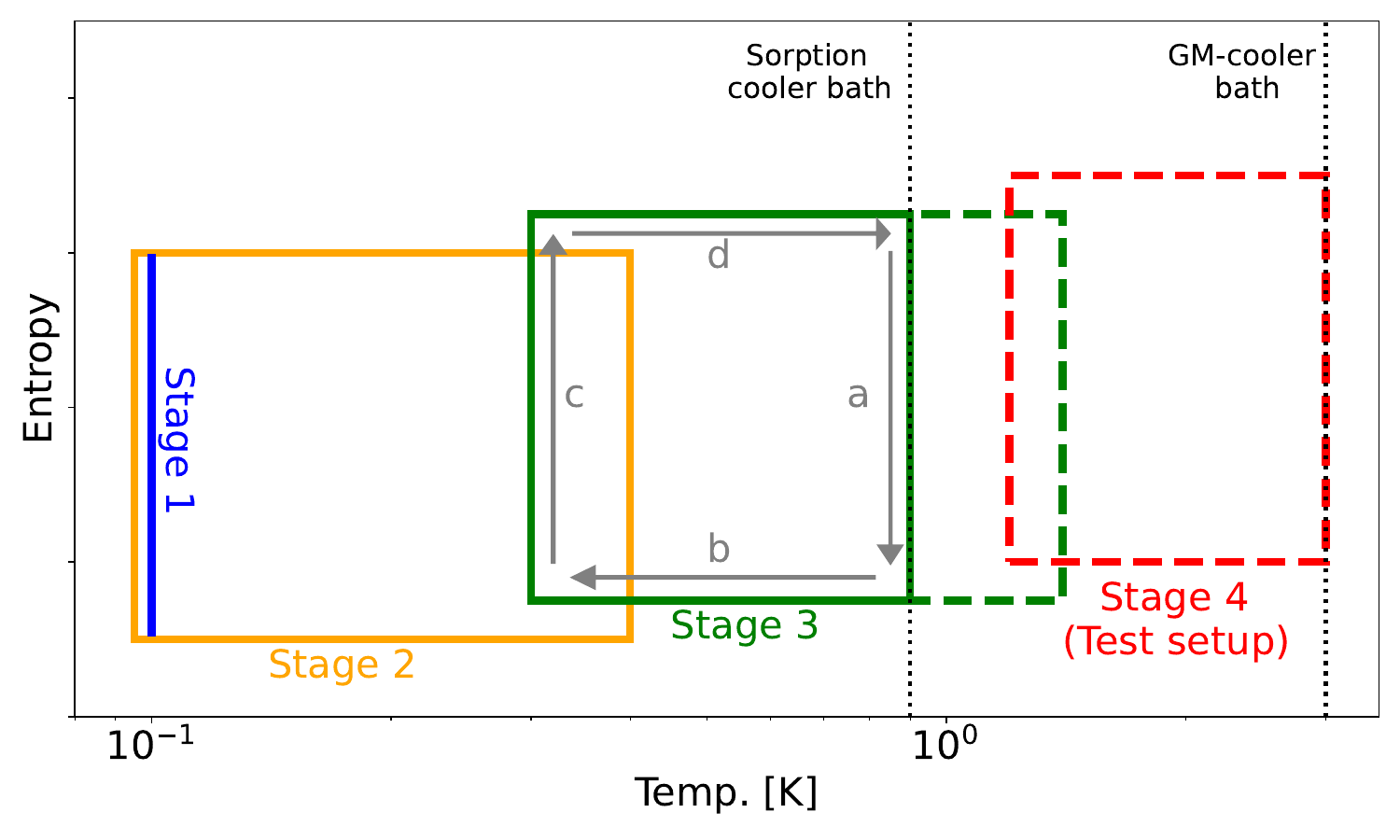}
   \end{tabular}
   \end{center}
   \caption
%>>>> use \label inside caption to get Fig. number with \ref{}
   { \label{fig:CADR} \textit{Left:} The three-stage EXCLAIM CADR assembled on the receiver mounting ring. \textit{Right:} Schematic of multi-stage CADR operation. While Stage 1 is continuously maintained at the base temperature, the upper stages cascade heat to the bath. The different legs of the thermodynamic cycles are described in Section \ref{sec:CADR}. The dashed lines show the cycles associated with a fourth stage used during the lab tests.}
   \end{figure}

EXCLAIM uses a three-stage CADR, shown in Figure \ref{fig:CADR}, to provide continuous cooling to the 100 mK stage thermally connected to the spectrometer package. Each CADR stage is comprised of a paramagnetic salt crystal surrounded by a superconducting magnet made from NbTi wire wound onto a mandrel \cite{shirron2004}. While the salt pills for the two warmer stages (S2 and S3) are suspended within the mandrel bore by Kevlar assemblies with minimal thermal conduction, the coldest stage (S1) pill does not require the thermal suspension from its magnet due to low hysteresis heating. Based on the PIPER heritage\cite{switzer2019}, all three stages use chromium potassium alum (CPA) salt as the refrigerant for its relatively high cooling power at low operating temperatures, non-corrosive property, and ease of crystal growth \cite{shirron2014}. Passive gas-gap heat switches\cite{kimball2017}, consisting of a thin titanium chamber filled with $^3$He and gold-plated copper fins, thermally connect S2 to S3 and S3 to the sorption cooler. As the cold end of the heat switch drops below a target temperature determined by the $^3$He fill pressure, $^3$He adsorbs onto the fins, thermally isolating the two ends. The S2 and S1 stages are connected with an active superconducting heat switch (SCHS), where a thermal connection is established by applying current to a small Helmholtz coil that drives an interconnecting superconducting lead normal. The SCHS is used between S1 and S2 because $^3$He does not have sufficient vapor pressure to act as a thermal conductor below $\sim$ 250 mK. It also enables rapid control of the S2--S1 thermal exchange, providing flexibility in maintaining the continuous S1 temperature.

The operating concept of the EXCLAIM three-stage CADR is shown in the schematic in Figure \ref{fig:CADR}. During the continuous operation, S2 and S3 perform conceptually the same thermodynamic cycle. In \textit{leg a}, the stage isothermally magnetizes against an exchange at $T_\mathrm{recycle}$, reducing the entropy of the magnetic spins. In \textit{leg b}, the stage thermally decouples from the exchange and adiabatically demagnetizes, cooling the salt pill to $T_\mathrm{cold}$ (below $T_\mathrm{recycle}$ of the lower stage). In \textit{leg c}, it isothermally absorbs heat from the lower stage until it is out of cooling power or the lower stage is fully charged. Finally, in \textit{leg d}, it thermally decouples from the lower stage and adiabatically magnetizes, bringing the salt pill back to $T_\mathrm{recycle}$. During the continuous operation, S1 is servoed at the 100\,mK base temperature, and the active SCHS is used to extract heat from S1 when S2 gets colder. Table~\ref{tab:CADR_param} summarizes the properties of the EXCLAIM CADR stages, including the inductance ($L$), maximum current ($I_\mathrm{max}$), magnetic field-to-current ratio ($B/I$), and the $T_\mathrm{recycle}$ and $T_\mathrm{cold}$ values from the lab test setup with a fourth stage (see Section \ref{sec:cadr test}).

\begin{table}[h]
\begin{center} 
\begin{threeparttable}
\caption{Properties of the EXCLAIM CADR stages} 
\label{tab:CADR_param}      
\begin{tabular}{lccccccc}
\hline
\hline
Stage & Salt & Mass & $L$ & $I_\mathrm{max}$ & $B/I$ & $T_\mathrm{recycle}$$^{\dagger}$ & $T_\mathrm{cold}$$^{\dagger}$\\
\hline
S4$^{*}$ & GLF & 50 g & 40.0 H & 2.7 A & 1.25 T/A & 3 K & 1.2 K \\
S3  & CPA & 100 g & 9.0 H & 4.0 A & 0.3 T/A & 1.4 K & 0.32 K \\
S2  & CPA & 100 g & 11.5 H & 2.5 A & 0.3 T/A & 0.42 K & 95 mK\\
S1  & CPA & 55 g & 0.47 H & 0.9 A & 0.1 T/A & - & 100 mK \\
\hline
\end{tabular}
\begin{tablenotes}
\small
\item $^{*}$Only used for the test setup
\item $^{\dagger}$Values used for the test setup and will be optimized separately for the flight
\end{tablenotes}
\end{threeparttable}
\end{center}
\end{table}

Each CADR stage has one primary and one redundant ruthenium oxide (ROX) thermometer read out through 76 $\mu$m CuNi-clad NbTi wires to limit parasitic conduction. The electrical interface to the CADR is provided through a custom printed circuit board (PCB), shown in Figure \ref{fig:CADR}, which brings in high-current lines from the receiver and provides voltage taps to measure the drop across the superconducting coil in a four-wire configuration. Outside the receiver, vapor-cooled high-current copper leads are directly soldered to the pins of a superfluid-tight feedthrough (see Figure \ref{fig:exclaim_CAD}). This feedthrough also carries lines for the sorption cooler pump heater and its gas-gap heat switch. Inside the receiver, we use 0.31 mm diameter NbTi wires with 0.5 mm copper cladding to transmit up to $\sim$ 4 A current (Table \ref{tab:CADR_param}) to the CADR. These NbTi wires are soldered to the feedthrough on one end and gold-plated copper bobbins on the other end. The bobbins are attached to the PCB with fasteners and spring washers. The voltage taps are carried out from an MDM31 connector on the PCB and follow the same wiring scheme used for CADR thermometry and elsewhere in the receiver.

\section{Lab Testing}
\label{sec:test}
Once the receiver is fully integrated, it will be tested in flight-like conditions in a smaller LHe dewar (48.6-cm diameter and 152-cm depth) in the lab and transferred to the telescope with no changes in configuration for the flight. However, the duration of this integrated receiver test is limited by the availability of LHe, requiring $\sim$~100~L of LHe per day of testing. Therefore, we have independently tested and qualified major receiver components before their integration into the receiver.

\subsection{Window Test}
\label{sec:window_test}
\begin{figure}[h]
   \begin{center}
   \includegraphics[height=7cm]{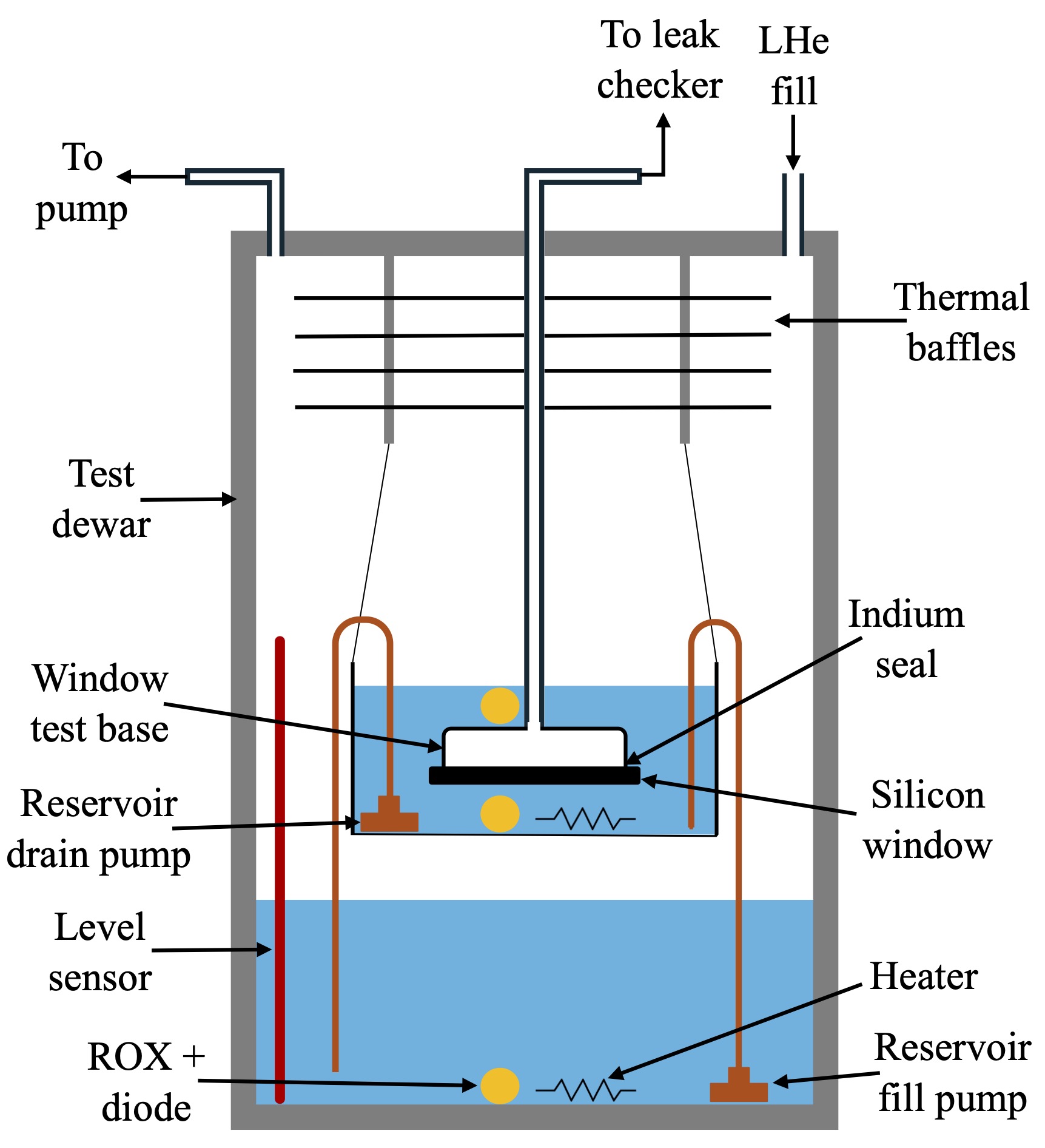}
   \end{center}
   \caption{ \label{fig:wndw_test_schm} Schematic of the setup for testing the superfluid-tight window seal. We cooled down a copy of the flight window in a LHe test dewar and measured the helium leak rate in the test volume.}
   \end{figure}

\begin{figure}[b]
   \begin{center}
   \includegraphics[height=5cm]{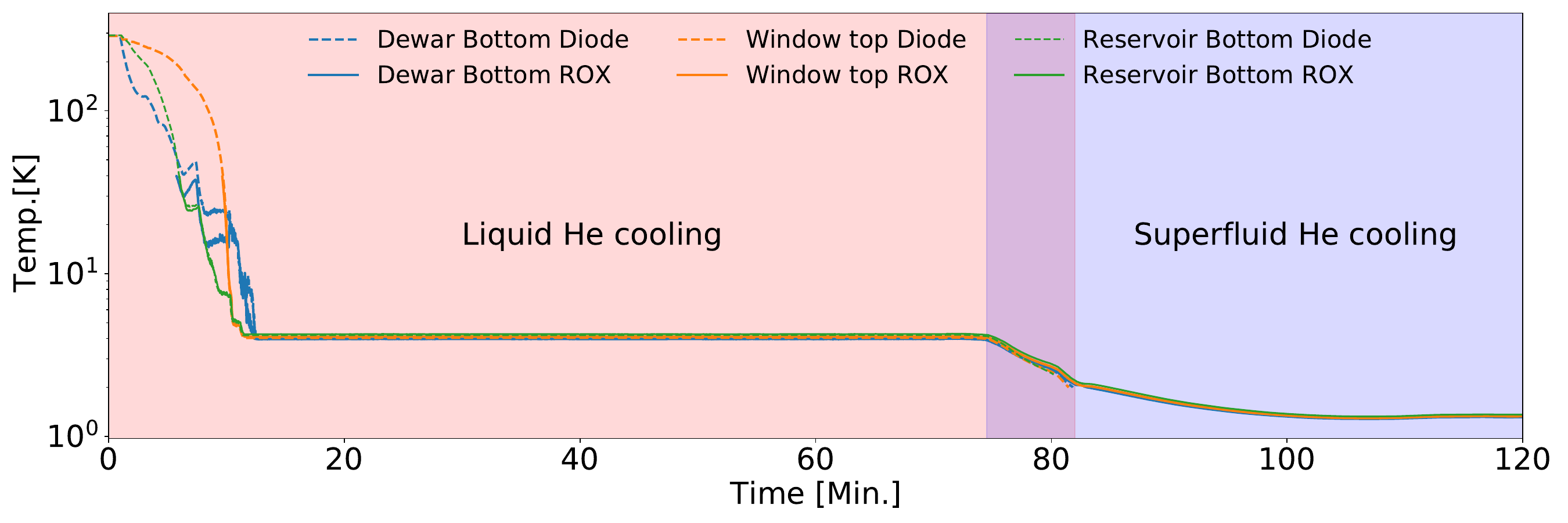}
   \end{center}
   \caption{ \label{fig:wndw_test_result} Temperature plot of various locations (yellow dots in Figure \ref{fig:wndw_test_schm}) in the window test setup. The test volume maintained $2 \times 10^{-3}$ mbar pressure with a background He leak rate of $5\times 10^{-8}$ mbar.l/s throughout the cool down, demonstrating a superfluid-tight seal.}
   \end{figure}

As described in Section \ref{sec:intro}, all receiver interfaces must remain superfluid tight, including the interface between the receiver lid and the silicon window. To our knowledge, EXCLAIM is the first instrument to employ a silicon vacuum window in the receiver with stringent requirements for superfluid tightness. We, therefore, test the ability of the indium seal to create a superfluid-tight interface between the receiver lid and the window. We mount a copy of the flight window onto a test baseplate with the same interface as the flight receiver lid on one side and a superfluid-tight vacuum interface to connect to a leak checker pump on the other side. The schematic of the test setup is shown in Figure \ref{fig:wndw_test_schm}. Following a similar procedure to that described in Ref. \citenum{datta2021}, the indium seal is compressed using titanium fasteners with a stack of spring washers on a retainer ring. The test window hangs inside a small reservoir supplied with superfluid helium pumps, as shown in Figure \ref{fig:wndw_test_schm}. With the leak checker continuously pumping on the test volume behind the window, we first pour LHe into the test dewar and then pump on the system to create flight-like conditions. As indicated by the $\sim$ 1.4 K temperature of the ROX on top of the window test base, as shown in Figure \ref{fig:wndw_test_result}, we were able to submerge the window along with the indium seal in superfluid helium. Throughout the cool down, the test volume maintained $2 \times 10^{-3}$\,mbar pressure with a background He leak rate of $5\times 10^{-8}$ mbar~l/s, demonstrating a superfluid-tight seal. We plan to repeat this test with the AR-coated flight window and similarly test the high-current feedthrough interface.

\subsection{Sorption-cooler Test}
\label{sec:sorption_test}
We mounted the EXCLAIM sorption cooler on a 3-K bath provided by a pulse-tube cryocooler in the lab to verify its operation and performance. We followed the operational procedure described in Section \ref{sec:sorption}. After cycling the cooler, the cold head reached 850 mK without any external load. Since the CADR S3 rejects heat to the sorption cooler, the loading on the sorption cooler during the flight depends on the frequency of S3 recycles. For nominal CADR operation at float, we estimate 82 $\mu$W of loading on the sorption cooler \cite{switzer2021}. To test its cooling capacity, we supplied 100 $\mu$W of power (18 $\mu$W higher than the expected load) using a 2 k$\Omega$ heater on the cold head. During the initial testing, the cold head was able to maintain $\sim$ 890 mK for $\gtrsim$35 h with 100 $\mu$W load. While this measured cooling capacity meets the baseline cold operation requirements for the flight, we plan to further optimize the operational procedure since the duration over which the head remains cold depends on the efficiency with which the helium charge is initially condensed \cite{chase2020}. The integrated receiver test will also help us optimize the frequency of the S3 recycles needed during the flight to better understand the loading on the sorption cooler.   

\subsection{CADR Test}
\label{sec:cadr test}

\begin{figure}[b]
\begin{center}
\includegraphics[height= 8.0cm]{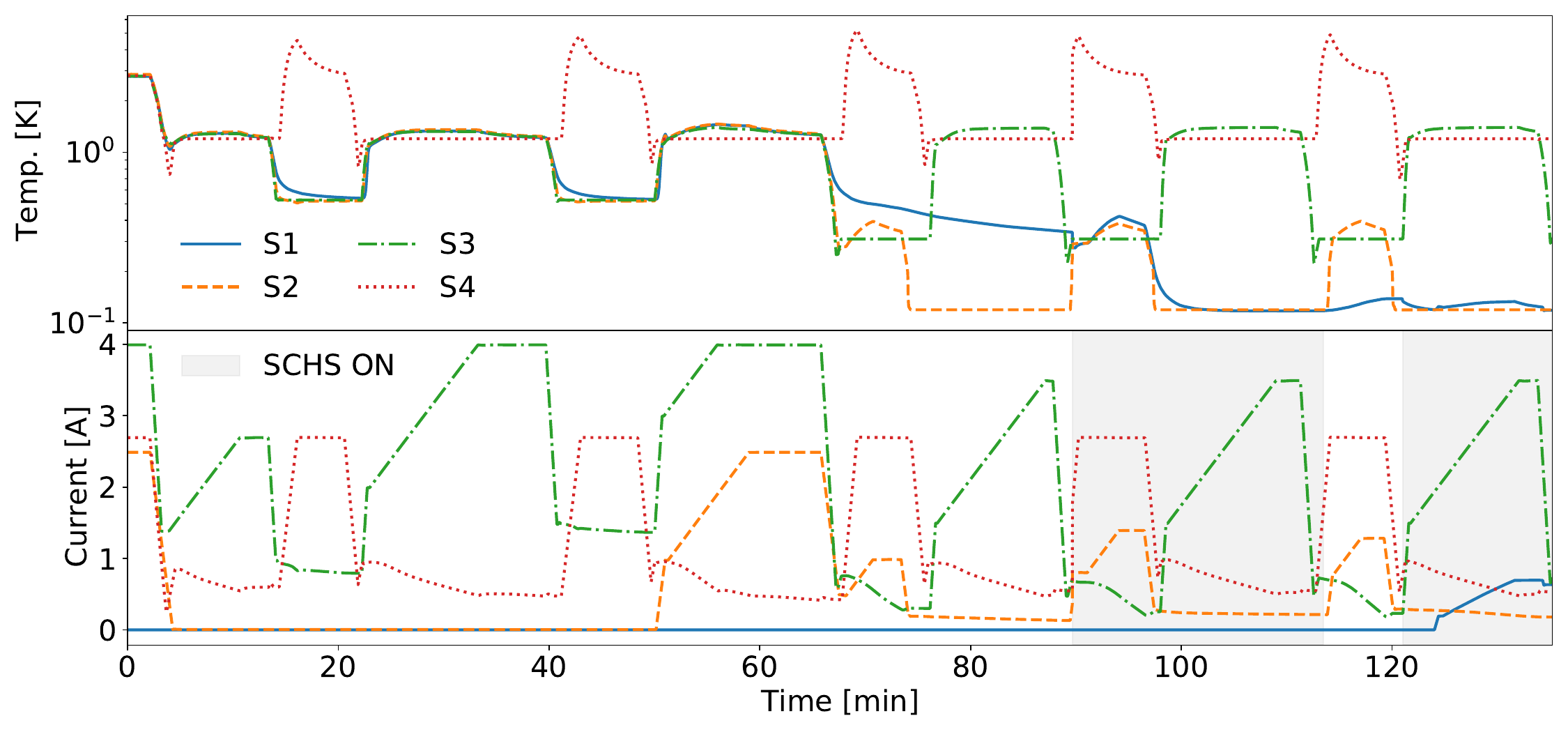}
\caption{ \label{fig:cadr_init} The temperature (\textit{top}) and current (\textit{bottom}) data showing the CADR initialization steps for the test setup with a 3-K bath in preparation for the continuous operation shown in Figure \ref{fig:cadr_cont}. The plots highlight the bootstrapping technique used to recycle S2 and S3 until they have sufficient capacity to cool S1 down to its operating temperature.}
\end{center}
\end{figure}
\begin{figure}[h]
\begin{center}
\includegraphics[height= 10cm]{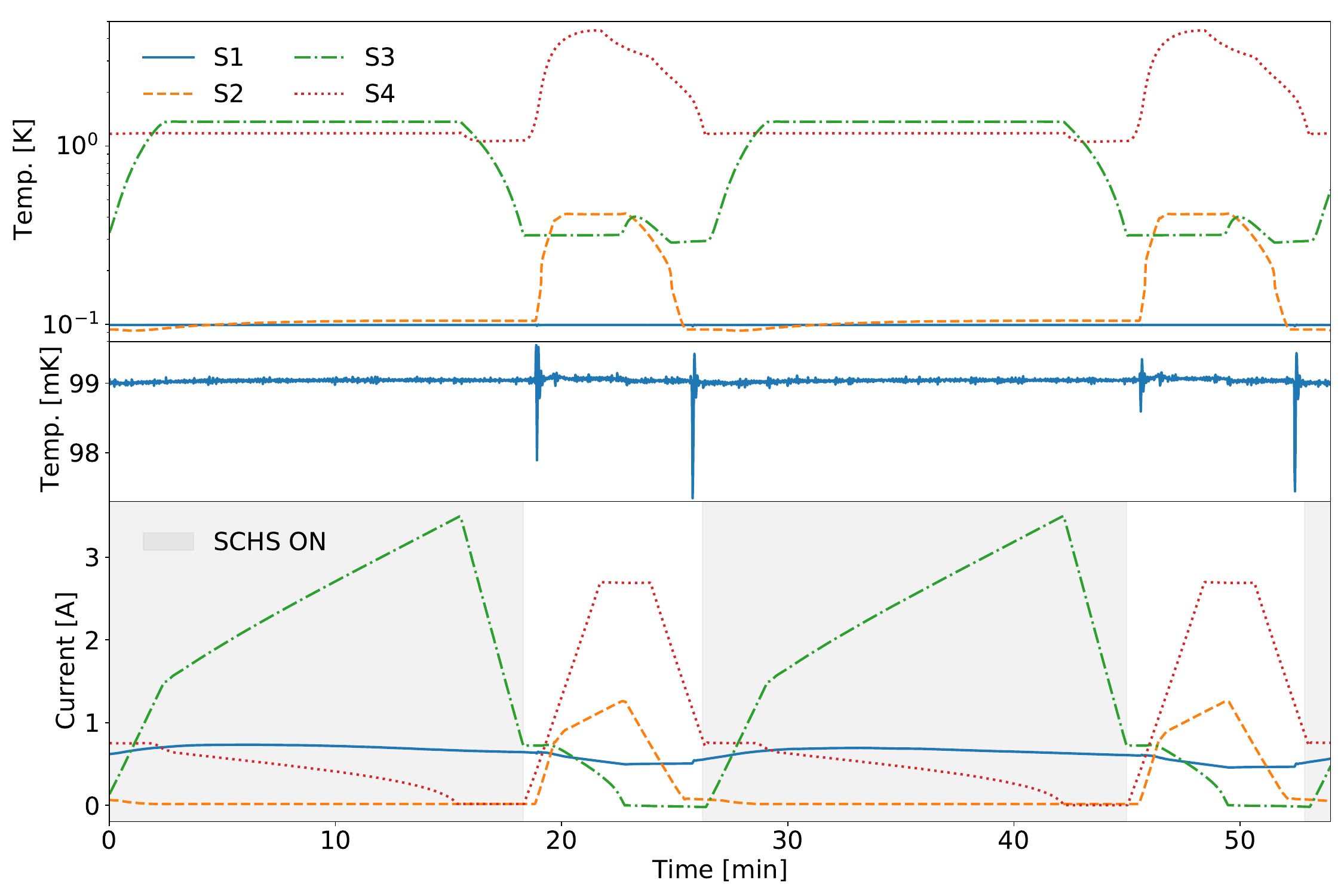}
\caption{ \label{fig:cadr_cont} The temperature (\textit{top}) and current (\textit{bottom}) data showing two typical cycles from the continuous operation of the CADR in the lab test setup. While S1 is maintained at 99 mK, heat is cycled out through other stages to the bath. The zoomed-in plot in the \textit{middle} highlights the stability of the S1 temperature during the operation.}
\end{center}
\end{figure}

As seen in Figure \ref{fig:CADR}, the EXCLAIM CADR has now been assembled on its mounting ring and is ready to be integrated into the receiver. As other receiver parts needed to integrate the receiver core are being fabricated, we tested the three-stage EXCLAIM CADR using a gadolinium lithium fluoride (GLF) Stage 4 (S4) instead of the sorption cooler. For this test, shown in Figures \ref{fig:cadr_init} and \ref{fig:cadr_cont}, S4 was thermally connected to a 3 K heat sink provided by a Gifford-McMahon (GM) cryocooler via a passive gas-gap heat switch. In addition to testing and verifying the CADR performance, this test was also used to qualify the EXCLAIM flight electronics and software for controlling and automating the CADR operation. 

In the test setup, the S4 -- bath passive heat switch is thermally conductive down to $\sim$ 1.7 K. With this condition, even when all four stages are fully charged to their respective $I_\mathrm{max}$ values (Table \ref{tab:CADR_param}), the CADR does not have sufficient cooling capacity to pull S1 down to its operating temperature from the 3-K bath with a single demagnetization step. Instead, we use a ``bootstrapping'' technique to build the cooling capacity of the warmer stages over multiple cycles, as shown in Figure \ref{fig:cadr_init}. During the first demagnetization step, as S4 approaches 1.2~K, we start re-building current in S3 while keeping S4 at 1.2~K through a proportional-integral-derivative (PID) control loop in the flight electronics. At this point, the S4 -- bath heat switch is non-conducting, allowing S3 to magnetize isothermally. When S4 runs out of cooling power, we ramp down S3 to $\sim$ 0.5 K (making the S3 -- S4 heat switch non-conducting) and let S4 recycle. Following a similar procedure, we build current in S2 as shown in Figure \ref{fig:cadr_init}. Once S3 and S2 have built sufficient current, the SCHS is turned ON, letting S2 cool S1 down to $\sim$~100~mK. Finally, we magnetize S1 in preparation for the continuous operation. While this CADR initialization algorithm is time-consuming ($\sim$ 2 hours), it enables prolonged testing of the continuous operation (Figure \ref{fig:cadr_cont}) from a 3-K bath with a simple passive heat switch. For the flight, the CADR will be operating from a 900-mK bath, significantly reducing the initialization time, which we plan to test and optimize during the integrated receiver test. 

Once the CADR has been initialized with S1 at $\sim$ 100 mK temperature and $\sim$ 0.8 A current, the continuous operation (described conceptually in Figure \ref{fig:CADR} and Section \ref{sec:CADR}) can begin. Figure \ref{fig:cadr_cont} shows two typical cycles of continuous operation from the lab test with S1 temperature stable at 99 mK, showing variations below $\sim$~1~mK level. The S2, S3, and S4 cyclic operations are autonomously controlled \cite{switzer2019} from the flight electronics, thus requiring no user input from the ground during the flight. The S1 is PID-controlled at the operating temperature, and the SCHS is manually controlled to allow the operator to recharge S1 as needed. During the lab test, we also qualified the CADR operation with various heater loads on S1 to simulate the parasitic loading during the flight. The CADR initialization and continuous operation, including the S1 temperature stability, will be further optimized during the integrated receiver test with the sorption cooler in place.

\section{summary}
\label{sec:summary}
EXCLAIM is a balloon-borne mission designed to survey star formation over cosmological time scales using intensity mapping. It uses an open LHe bucket dewar to house a fully cryogenic telescope that allows fast integration in dark atmospheric windows. Since the receiver sits in a superfluid helium bath at float, all the receiver interfaces, including the silicon window, were designed to be superfluid-tight. The receiver houses a three-stage CADR that maintains the spectrometer package at 100 mK and a sorption cooler that provides a 900\,mK intermediate thermal stage. Given the limited test time with LHe for the integrated receiver, we have independently tested and qualified major receiver components. Next, we plan to integrate the receiver core and test it in a custom pulse-tube-cooled system, followed by a LHe dewar test before the planned flight in September 2025.

\acknowledgments % equivalent to \section*{ACKNOWLEDGMENTS}       
 EXCLAIM began in April 2019 as a 5-year NASA Astrophysics Research and Analysis (APRA 1263 17-APRA17-0077) grant. S.D. is supported under NASA-JHU Cooperative Agreement 80NSSC19M005.

% References
\bibliography{report} % bibliography data in report.bib
\bibliographystyle{spiebib} % makes bibtex use spiebib.bst

\end{document}